\documentstyle[psfig]{aipproc}

\begin{document}

\psfull

\title{Increased Sensitivity to Possible
Muonium to Antimuonium Conversion}

\author{V. Meyer$^1$, A. Grossmann$^1$, K. Jungmann$^1$, J. Merkel$^1$,
G. zu Putlitz$^1$, I. Reinhard$^1$, K. Tr\"ager$^1$, P.V. Schmidt$^1$,
L. Willmann$^1$,
R . Engfer$^2$, H.P. Wirtz$^2$,
R. Abela$^3$, W. Bertl$^3$, D. Renker$^3$, H.K. Walter$^3$,
V. Karpuchin$^4$, I. Kisel$^4$, A. Korenchenko$^4$, S. Korenchenko$^4$,
N. Kravchuk$^4$, N. Kuchinsky$^4$, A. Moiseenko$^4$,
J. Bagaturia$^5$, D. Mzavia$^5$, T. Sakhelashvili$^5$,
V.W. Hughes$^6$}
\address{
 $^1$Physikalisches Institut, Universit\"at Heidelberg, D-69120 Heidelberg, 
 Germany;
 $^2$Physik Institut, Universit\"at Z\"urich, CH-8057 Z\"urich, Switzerland;
 $^3$Paul Scherrer Institut, CH-5232 Villigen, Switzerland;
 $^4$Joint Institute of Nuclear Research, RU-141980 Dubna, Russia;
 $^5$Tbilisi State University, GUS-380086 Tbilisi, Georgia;
 $^6$Physics Department, Yale University, New Haven CT 06520, USA}

\maketitle

\begin{abstract}
A new experimental search for muonium-antimuonium conversion was
conducted at the Paul Scherrer Institute, Villigen, Switzerland. 
The preliminary analysis yielded one event fulfilling all required 
criteria at an expected background of 1.7(2) events due to accidental
coincidences. An upper limit for the conversion probability in 0.1 T
magnetic field is extracted as $8 \cdot 10^{-11}$ (90\%~CL).
\end{abstract}


The hydrogen like muonium atom ($M= \mu^+ e^-$) consists of two leptons from
different generations.
The close confinement of the bound state offers excellent opportunities to
explore precisely fundamental electron-muon interaction. The dominant
part of the binding in this system is
electromagnetic and can be calculated to very high
accuracy in the framework of quantum electromagnetics (QED). Indeed, precision
experiments on electromagnetic transitions in muonium have been employed
both to verify bound state QED calculations and for determining
most accurate values of fundamental constants \cite{Hughes_90}.
                                                               
Since the effects of all known fundamental forces in muonium are calculable
very well, it renders the possibility to search sensitively for yet unknown
interactions between both particles. A conversion of muonium into
its antiatom ($\overline{M} = \mu^- e^+$) would violate additive lepton
family number conservation and is not provided in standard theory.
However, muonium-antimuonium conversion appears
to be natural in many speculative
theories, which try to extend the standard model in order to explain
some of its yet not well understood features like parity violation in
weak interaction and particle mass spectra.
The interaction could be mediated by a doubly charged 
Higgs boson \protect{\cite{Herczeg_92}},
heavy Majorana neutrinos \protect{\cite{Halprin_82}},
a neutral scalar \protect{\cite{Hou_96}}, e.g.
a supersymmetric $\tau$-sneutrino
\protect{\cite{Mohapatra_92}} or a dileptonic gauge boson 
\protect{\cite{Sasaki_94}}.

An experiment had been set up to search for spontaneous
muonium-antimuonium conversion at the Paul Scherrer Institute (PSI)
in Villigen, Switzerland \cite{Abela_96}.
It uses the powerful signature developed
in an experiment at the Los Alamos Meson Physics Facility (LAMPF) USA, 
which requires the coincident identification
of both constituents of the antiatom in its decay \cite{Matthias_91}.

Muonium atoms were produced by stopping a beam of surface muons
in a SiO$_2$ powder target, where a fraction of them forms
muonium by electron capture, some of which diffuse through the target 
surface with thermal energies into vacuum. 
Energetic electrons from the decay of
the $\mu^-$ in the antiatom can be observed in a magnetic spectrometer
at 0.1 T magnetic field
consisting of five concentric multiwire proportional chambers
and a 64 fold segmented hodoscope. 
The positron in the atomic shell of the antiatom
is left behind after the decay with 13.5 eV average kinetic energy.
It can be electrostatically accelerated to 8 keV and guided
in a magnetic transport system onto a position sensitive microchannel
plate detector (MCP). Annihilation radiation can be observed in a 12 fold
segmented pure CsI calorimeter surrounding the MCP.

The muonium production was monitored regularly
by reversing all electric and magnetic fields of the instrument
every five hours for a duration of 20 minutes.
Targets had to be replaced twice a week because of observed
deterioration of muonium production on a one week time scale.
In the course of the experiment $5.7 \cdot 10^{10}$ muonium atoms
were observed in the interaction volume for antimuonium decays.
There was one event which passed all required  criteria, i.e.
fell into a 99\% confidence interval of each relevant distribution.
The expected background due to accidental coincidences is 1.7(2) events.

The preliminary combination of all data
recorded in the experiment between 1993 and 1996 
\cite{Abela_96,Schmidt_97,Wirtz_97}
results in
an upper limit
for the conversion probability in 0.1 T magnetic field of
${\rm P}_{{\rm M} \overline{{\rm M}}}\leq 8 \cdot 10^{-11}$ (90~\%~CL).
For an assumed effective (V-A)x(V-A) type four fermion interaction
this corresponds to an upper limit
for the coupling constant of
${\rm G}_{{\rm M} \overline{{\rm M}}}\leq 3\cdot 10^{-3} {\rm G}_{\rm F}$(90~\%~CL),
where ${\rm G}_{\rm F}$
is the weak interaction Fermi coupling constant \cite{Schmidt_97}.

This new result allows to rule out definitively
a certain  ${\rm Z}_8$ model with more than three particle generations 
\cite{Hou_96}
and to set a new lower limit of 2.6 TeV/c$^2$ * $g_{3l}$ 
($g_{3l}$ depends on model, of order unity)
on the mass of a dileptonic gauge boson in GUT models - well
beyond the value extracted from high energy Bhabha scattering \cite{Sasaki_94}.
It can be further shown in the framework of minimal left right symmetric and
supersymmetric models \cite{Herczeg_97} that lepton number violating muon decay
($\mu^+ \rightarrow e^+ + \nu_{\mu} + \overline{\nu}_e$) is not an option
for explaining the excess neutrino counts in the LSND neutrino
experiment at Los Alamos \cite{LSND_96}.

\begin{figure}[t]
\label{theo_mmb}
\unitlength 1.0 cm
 \begin{minipage}{5.0cm}
 \begin{picture}(5.0,5.7)
  \centering{
   \hspace{-1.9 cm}
   \raisebox{0.0 cm}{
   \mbox{
   \psfig{figure=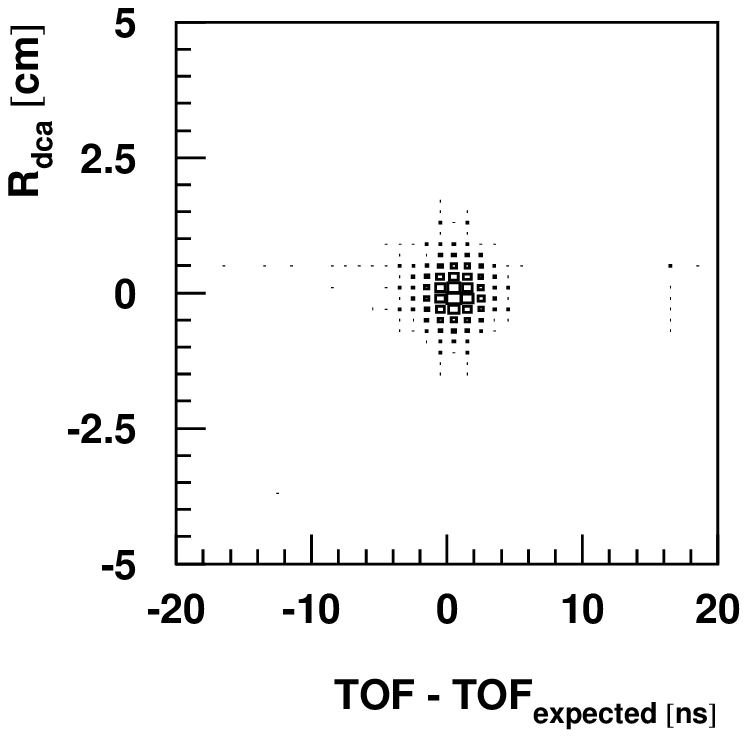,height=5.7cm}
        }
        }
             }
 \end{picture}
 \end{minipage}
 \hspace{0.3cm}
 \begin{minipage}{5.0cm}
 \begin{picture}(5.0,5.7)
  \centering{
   \raisebox{0.0 cm}{
   \mbox{
   \psfig{figure=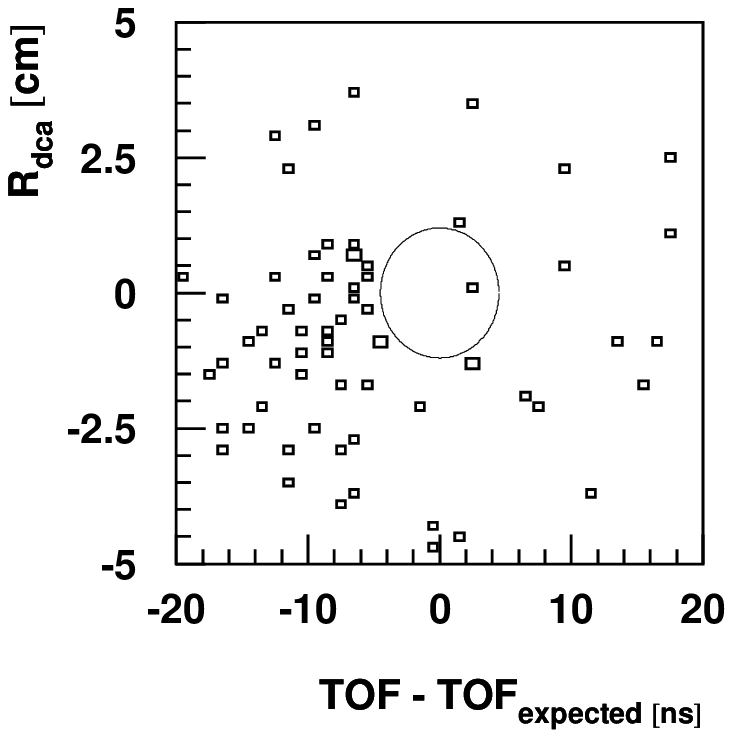,height=5.7cm}
        }
        }
             }
 \end{picture}
 \end{minipage}
 \centering\caption[]
        {
        The distribution of the distance of closest approach (R$_{dca}$) 
        between a track from an energetic 
        particle in the magnetic spectrometer and the back projection of the
        position on the MCP detector versus the time of flight (TOF)
       of the atomic shell particle for a muonium measurement (left)
        and for all data recorded in 1996 while searching for 
        antimuonium (right). 
        One single event falls within 3 standard deviations region
        of the expected TOF and R$_{dca}$ which is indicated by the ellipse.
         The events concentrated at early times 
       and low R$_{dca}$correspond 
        to a background signal from the allowed decay
        $\mu \rightarrow 3e+2\nu$. 
       }
\end{figure}

This work was supported by the Bundesminister f\"ur Bildung und Forschung (BMBF)
of Germany, the Schweizer Nationalfond, the Russian Federation for Fundamental
Research and a NATO research grant.

\end{document}